\documentclass[apj]{emulateapj}
\usepackage{apjfonts}

\def\arcsec{$^{\prime\prime}$}

\shorttitle{A giant radio halo in a new Planck cluster}
\shortauthors{S.~Giacintucci et al.}

\begin{document}

\setlength{\pdfpageheight}{\paperheight}
\setlength{\pdfpagewidth}{\paperwidth}

\slugcomment{Received 2012 November 7; accepted 2013 January 31; published 2013 March 1}

\title{DISCOVERY OF A GIANT RADIO HALO IN A NEW PLANCK GALAXY CLUSTER PLCKG171.9-40.7}

\author{Simona Giacintucci\altaffilmark{1,2}, 
Ruta Kale\altaffilmark{3,4},
Daniel R. Wik\altaffilmark{5,6},
Tiziana Venturi\altaffilmark{3},
Maxim Markevitch\altaffilmark{5}
}
\altaffiltext{1}{Department of Astronomy, University of Maryland,
  College Park, MD 20742, USA; simona@astro.umd.edu}
\altaffiltext{2}{Joint Space-Science Institute, University of Maryland, College Park,
MD, 20742-2421, USA}
\altaffiltext{3}{INAF - Istituto di Radioastronomia, via Gobetti 101, I-40129 Bologna,
Italy}
\altaffiltext{4}{Dipartimento di Fisica e Astronomia, via Ranzani 1,
  I-40127 Bologna, Italy}
\altaffiltext{5}{Astrophysics Science Division, NASA/Goddard Space
  Flight Center, Greenbelt, MD 20771, USA}
\altaffiltext{5}{NASA Postdoctoral Position Fellow}

\begin{abstract}
We report the discovery of a giant radio halo in a new, hot, X-ray
luminous galaxy cluster recently found by {\em Planck}, PLCKG171.9-40.7.
The radio halo was found using Giant Metrewave Radio Telescope
observations at 235 MHz and 610 MHz, and in the 1.4 GHz data from a NRAO 
Very Large Array Sky Survey pointing that we have reanalyzed. 
The diffuse radio emission is coincident with the cluster X-ray emission, and
has an extent of $\sim$1 Mpc and a radio power of $\sim 5\times10^{24}$ W Hz$^{-1}$ at 1.4 GHz. Its integrated radio spectrum has a slope of $\alpha\approx1.8$
between 235 MHz and 1.4 GHz, steeper than that of a typical giant halo.
The analysis of the archival {\em XMM-Newton} X--ray data shows that 
the cluster is hot ($\sim 10$ keV) and disturbed, consistent with
X-ray selected clusters hosting radio halos. This is the first giant
radio halo discovered in one of the new clusters found by {\em Planck}.
\end{abstract}

\keywords{galaxies: clusters: general --- galaxies: clusters: individual
  (PLCKG171.9-40.7) --- intergalactic medium --- radio continuum:
  general --- X--rays:galaxies: clusters}

\section{Introduction}
\label{sec:intro}

Spectacular diffuse radio sources, with very low surface brightness 
and $\sim$Mpc extent, are observed in some of the most massive galaxy 
clusters. Depending on their location, i.e., cluster center versus
outskirts, these cluster-wide sources are
classified as giant radio halos or
relics \citep[see][for reviews; see
\cite{2012MNRAS.425L..36V,2011PhDT........14V,2010ApJ...718..939K,2012ApJ...744...46K,2011A&A...534A..57G,2011MNRAS.412....2B,
2012MNRAS.426...40B}
for recent results]{2008SSRv..134...93F, 2011MmSAI..82..499V,2012A&ARv..20...54F}.
Halos and relics are due to synchrotron emission from 
ultrarelativistic electrons with energy of several GeV,
spinning in $\mu$G magnetic fields that permeate the hot intracluster
medium (ICM).

Particle acceleration and  magnetic field amplification during 
cluster mergers have been proposed as mechanisms by which 
diffuse radio emission is generated in clusters. 
In particular, relics should be produced by electron (re)acceleration at 
merger shocks \citep{1998A&A...332..395E,2005ApJ...627..733M,
2007MNRAS.375...77H,2011ApJ...734...18K,2012ApJ...756...97K},
while giant halos 
may be caused by reacceleration of  
lower energy relativistic electrons by MHD turbulence 
during mergers \citep[][and references therein]{2011JApA...32..437B}.
Although alternative possibilities have been proposed for the origin
of the radio-emitting electrons  -- the secondary models 
\citep[e.g.,][]{1980ApJ...239L..93D,1999APh....12..169B,2004JKAS...37..455P,2010ApJ...722..737K,2011A&A...527A..99E} --
turbulent reacceleration models seem to be favored by current 
radio observations
\citep[e.g.,][]{2008A&A...484..327V,2008Natur.455..944B,2009A&A...507..661B,2010A&A...517A..43M,2010MNRAS.407.1565D,2011MNRAS.412....2B,2012MNRAS.426..956B}.
In particular, radio halos with very steep spectral index\footnote{We adopt the
convention $S_{\nu} \propto \nu^{-\alpha}$, where $S_{\nu}$ is the 
is the flux density at the frequency $\nu$.} ($\alpha >1.5$) have been
found \citep[the prototype case is A\,521 with $\alpha\sim2$,][]{2008Natur.455..944B,2009ApJ...699.1288D},
consistent with turbulent reacceleration in weak mergers, 
i.e., collisions between clusters with relatively low mass
($<10^{15}$ M$_{\odot}$) or accretion of smaller systems into a
massive cluster. Secondary models have difficulty explaining such 
ultra-steep spectrum radio halos (USSRHs), due to the large energy
required in the form of relativistic protons \citep[e.g.,][]{2008Natur.455..944B,2010A&A...517A..43M,2011A&A...533A..35V}. 
However, only two giant radio halos are 
confirmed USSRHs so far -- A\,521 and A\,697 -- based on a well-constrained spectrum with five data points between 150 MHz and
1.4 GHz (Macario et al. 2013). Other possible USSRHs still lack 
confirmation of their spectral index (e.g., Giacintucci et al. 2011) 
and are currently under investigation. Therefore, it is necessary to
confirm the candidates and find more of these USSRHs, with an 
accurate determination of their spectral index, before 
ruling out the secondary models.

A number of dedicated searches for cluster diffuse radio sources  
have been carried out in the past decade
\citep[e.g.,][]{1999NewA....4..141G,2001ApJ...548..639K,2007A&A...463..937V,2008A&A...484..327V,2009A&A...507.1257G,2009A&A...508...75V,2011A&A...533A..35V}, 
and the number of giant halos and relics has rapidly increased. 
While $\sim$50 objects are presently known (including
candidates), because of their low surface brightness, 
most of them lack detailed spectral information,
crucial to test theoretical models and understand the origin of 
the radio-emitting electrons \citep[e.g.,][]{2012arXiv1210.7617V}.

\par The {\em Planck}\footnote{{\em Planck} 
(http://www.esa.int/Planck), a project of the European Space Agency
(ESA) with instruments provided by two scientific consortia funded by ESA
member states (in particular the lead countries France and Italy), with
contributions from NASA (USA) and telescope reflectors provided by a
collaboration between ESA and a scientific consortium led and funded by
Denmark.} Early Sunyaev-Zel'dovich (ESZ) cluster sample \citep{2011A&A...536A...8P}
is an all-sky sample of clusters, detected by their multi-frequency
signature in the {\em Planck} microwave 
observing bands \citep[30--857 GHz;][]{2011A&A...536A...1P}.
Clusters are detected through the measurement of the spectral distortion of the 
cosmic microwave background (CMB) due to inverse Compton 
scattering of CMB photons by the thermal electrons in the ICM -- the 
Sunyaev Zel'dovich (SZ) effect \citep{1972CoASP...4..173S}.
The Compton $y$-parameter, a measure of the SZ effect, does not suffer 
dimming due to distance, and its integral over the cluster is a
measure of the total thermal energy in the cluster. 
Therefore, SZ surveys are a powerful tool to find
massive clusters at high redshifts. A total of 51 new clusters
discovered by {\em Planck} have been confirmed 
by follow-up X-ray {\em XMM-Newton} and optical observations 
\citep[][2012b]{2011A&A...536A...9P, 2012A&A...543A.102P}. Most
of these new clusters appear to be hot, massive systems with highly irregular and
disturbed X-ray morphologies, and thus good targets to search for new
radio halos and relics. Recently, double 
radio relics in one of the new {\em Planck} clusters, PLCKESZ G$287.0+32.9$,
have been
discovered by \cite{2011ApJ...736L...8B}
using the Giant Metrewave
  Radio Telescope (GMRT). Here, we report a GMRT detection of a giant 
radio halo in the newly discovered {\em Planck}
cluster PLCKESZ G171.94-40.65 (hereafter PLCK171). The cluster, 
whose general properties are summarized in Table \ref{tab:PLCK171}, is
an X-ray luminous ($L_{\rm X} \sim 10^{45}$ erg s$^{-1}$) system at a
redshift of $z_{\rm Fe}=0.27\pm0.01$, as determined through the 
{\em XMM} Fe K line spectroscopy \citep{2011A&A...536A...9P}.
An optical photometric redshift of $z_{\rm opt}=0.31\pm0.03$, based on 29 cluster members, is 
reported by \cite{2012A&A...543A.102P}.

In this paper, we adopt $z=0.27$ and the $\Lambda$CDM cosmology 
with H$_0$=70 km s$^{-1}$ Mpc$^{-1}$,  $\Omega_m=0.3$ and 
$\Omega_{\Lambda}=0.7$. Under these assumptions, $1^{\prime \prime}$
corresponds to 4.135 kpc.

%
%
\begin{table}
\caption[]{General properties of PLCKESZ
G171.94-40.65}
\begin{center}
\begin{tabular}{lc}
\hline\noalign{\smallskip}
\hline\noalign{\smallskip}
RA$_{\rm J2000}$ (h m s) & 03 12 57.4 \\
DEC$_{\rm J2000}$ ($^{\circ}$ $^{\prime}$ 
                     $^{\prime\prime}$) & 08 22 10 \\
$z_{\rm Fe}$  & 0.27 \\
$L_{\rm X \, [0.1-2.4] \, kev}$ ($10^{45}$ erg s$^{-1}$) & 1.13\\
$kT$ (keV) & 10.65 \\
M$_{500}$ (M$_{\odot}$) & $1.09\times10^{15}$ \\
$D_{L}$ (Mpc) & 1364.5 \\
angular scale (kpc/ \arcsec) & 4.135\\ 
\noalign{\smallskip}
\hline\noalign{\smallskip}
\end{tabular}
\end{center}
Notes to Table \ref{tab:PLCK171} -- RA$_{\rm J2000}$ and DEC$_{\rm
  J2000}$ are the coordinates of the X-ray peak, $z$ is 
from the iron K line. The luminosity 
$L_{\rm X \, [0.1-2.4] \, kev}$, temperature $kT$ and mass M$_{500}$ are 
estimated within $R_{500}$, where $R_{500}$ is the radius
corresponding to a density contrast of 500 \citep{2011A&A...536A...9P}.
\label{tab:PLCK171}
\end{table}
%
%

\section{Radio observations} \label{sec:radioobs}

\subsection{GMRT observations}\label{sec:gmrt}

PLCK171 was observed with the GMRT at 235 MHz and 610 MHz in 
October 2011, as part of a project to search for diffuse radio
emission in 8 {\em Planck} clusters (project $21\_017$). The 
cluster was observed for $\sim$5 hours in dual-frequency 
mode, recording LL polarization at 235 MHz and RR polarization
at 610 MHz. 

The data were collected in spectral-line mode, using the GMRT 
software backend \citep[GSB;][]{2010ExA....28...25R} with a total observing 
bandwidth of 32 MHz at 610 MHz divided in 512 spectral channels. 
A bandwidth of 6 MHz was used at 235 MHz. Details on
these observations are summarized in Table \ref{tab:obs}, which
reports observing date, frequency and total bandwidth (columns 1, 2 and 3), total time on
source (column 4), usable time after data editing (column 5), full-width half 
maximum (FWHM) and position angle (PA) of the full array (column 6), 
rms level ($1\sigma$) at full resolution (column 7).

%
\begin{table*}[htbp!]
\caption[]{Details of the GMRT observations.}
\begin{center}
\footnotesize
\begin{tabular}{ccccccccc}
\hline\noalign{\smallskip}
\hline\noalign{\smallskip}
Observation & $\nu$ & $\Delta \nu$  & t$_{\rm tot}$ & t & FWHM,
p.a. & rms \\ 
date & (MHz)& (MHz) & (hours) & (hours) & ($^{\prime
     \prime}\times^{\prime \prime}$, $^{\circ}$)&  (mJy
   beam$^{-1}$) \\
\noalign{\smallskip}
\hline\noalign{\smallskip}
Oct 22, 2011 & \phantom{0}235 $^a$& 6  &  5  & $\sim$1 &15.4$\times$9.0, 64  
& 0.80 \\
Oct 22, 2011 & \phantom{0}610 $^a$& 32 &  5  & $\sim$1 & 5.5$\times$4.5, 0 &
0.12\\
\noalign{\smallskip}
\hline\noalign{\smallskip}
\end{tabular}
\end{center}
Notes to Table 2 -- $a$: observed in simultaneous 235 MHz/610
MHz mode.
\label{tab:obs}
\end{table*}
%
%

The data sets were calibrated and reduced using the NRAO 
Astronomical Image Processing System (AIPS) package. 
The data were initially inspected to identify and remove bad
channels, time intervals and visibilities with radio 
frequency interference (RFI). The data were found to be 
affected by severe phase instabilities caused by ionospheric 
scintillation during most of the observation,
leaving only $\sim$1 hour of usable time at both 610 MHz and 235 MHz. 
The data were then calibrated. The flux density scale was set
using 3C48 and 3C147 as amplitude calibrators and the 
\cite{1977A&A....61...99B} scale. The source 0323+055 was used as a
phase calibrator. The bandpass calibration was carried out using
the flux density calibrators. 
Few of the central channels free of RFI
were used to normalize the bandpass for each antenna. 
After the bandpass calibration, the central 420 channels were averaged to 
35 channels each 0.78 MHz wide to reduce the size of the data set at 610
MHz, and, at the same time, to minimize the bandwidth smearing effects within
the primary beam of the GMRT antenna. At 235 MHz, 72 channels
were averaged to 18 channels of 0.25 MHz width each.

After further careful editing in the averaged data, 
a number of phase-only self-calibration cycles and 
imaging were carried out to reduce residual phase variations
and improve the quality of the final images. Wide-field imaging 
was implemented at each step of the self-calibration process,
to account for the non-planar nature of the sky. The final images
were made using the multi-scale CLEAN implemented in IMAGR, which 
results in better imaging of extended sources compared to the traditional 
CLEAN \citep[e.g.,][for a
detailed discussion see Appendix A in \cite{2009AJ....137.4718G}]{2006AJ....131.2900C}.
We used $\delta$-functions as model components for the
unresolved features and circular Gaussians for the resolved ones,
with increasing width to progressively highlight the extended 
emission during the clean.
The rms sensitivity levels achieved in the images at 
full resolution is $\sim$0.12 mJy beam$^{-1}$ at 610 MHz 
and $\sim$0.80 mJy beam$^{-1}$ at 235 MHz (Table 2).
We also produced images with lower resolutions, down to 
$40^{\prime \prime}$, by tapering 
the $uv$ data using UVTAPER and ROBUST in IMAGR. The noise 
reached in the images at the lowest resolution is $\sim$0.3 mJy
beam$^{-1}$ at 610 MHz and $\sim2.3$ mJy beam$^{-1}$ at 235 MHz.
Residual amplitude errors should be within $\sim 5\%$ at 610 MHz 
\citep[e.g.,][]{2004ApJ...612..974C}; a calibration uncertainty of $\sim 20\%$ is
estimated at 235 MHz.

\subsection{VLA observations}\label{sec:vla}

We also reprocessed and analyzed Very Large Array (VLA) data at 
1.4 GHz from the NVSS \citep[NRAO VLA Sky Survey;][]{1998AJ....115.1693C}
pointing containing PLCK171 (project AC308).
The data were calibrated and reduced in AIPS. We used the 
standard Fourier transform deconvolution method to produce 
the images (CLEAN and RESTORE), and applied self-calibration to reduce the effects of 
residual phase errors in the data. The rms sensitivity level achieved
in the final image is $\sim$0.2 mJy beam$^{-1}$, with a restoring beam
of $58^{\prime\prime}\times48^{\prime\prime}$. Our image is $\sim2$
 times more sensitive than the public NVSS image, which has a local rms
 noise of $\sim$0.4 mJy beam$^{-1}$ ($50^{\prime\prime}$ restoring
beam). Residual amplitude errors are within $\sim5\%$.

\begin{table*}[t]
\caption[]{Discrete radio sources in PLCK171}
\begin{center}
\begin{tabular}{ccccccc}
\hline\noalign{\smallskip}
\hline\noalign{\smallskip}
Source& RA$_{\rm J2000}$ & DEC$_{\rm J2000}$ & $S_{\rm
  tot, \, 610 \, MHz}$  & $S_{\rm
  tot, \, 235 \, MHz}$& $\alpha_{\rm 610 \, MHz}^{\rm 235 \, MHz}$ & morphology \\
          & (h, m, s) & (${\circ}$, ${\prime}$, ${\prime \prime}$) &
          (mJy) &
 (mJy) & & \\
\noalign{\smallskip}
\hline\noalign{\smallskip}
S1  & 03 12 53.2 & +08 23 12 & $11.7\pm0.6$ & $>15.5\pm3.3$ &
$>0.3\pm0.2$ & NAT\\
S2  & 03 12 53.9 & +08 23 05 & $4.3\pm0.2$  &$14.6\pm3.1$ &
$1.3\pm0.2$ &unresolved\\
S3  & 03 12 56.9 & +08 22 11 & $2.7\pm0.2$ & $^{\dagger}$& - &unresolved\\
S4  & 03 12 57.5 & +08 22 09 & $85.8\pm4.3$ &$^{\dagger}$ & - & NAT\\
S5  & 03 12 57.3 & +08 21 37 & $6.0\pm0.3$ &$>8.1\pm1.9$ & 
$>0.3\pm0.3$ & NAT\\
S6  & 03 12 58.3 & +08 21 14 & $39.2\pm2.0$ & $66.5\pm13.3$ & $0.6\pm0.2$ & NAT\\
S7  & 03 12 54.4 & +08 20 35 & $1.9\pm0.1$ &- & - & unresolved \\
\hline
\end{tabular}
\end{center}
Notes to Table \ref{tab:flux} -- Peak coordinates are from the GMRT
full-resolution image at 610 MHz (Fig.~\ref{fig:610}). Flux densities have 
been measured on the GMRT full-resolution images at 610 MHz and 235
MHz (Fig.~\ref{fig:235u}). S1 and S5 are partially detected at
  235 MHz, therefore lower limits are reported for $S_{\rm
  tot, \, 235 \, MHz}$  and $\alpha$. $^{\dagger}$ Sources S3 and S4 are not resolved in
the 235 MHz image. Their combined flux density at 235 MHz is 248.4 mJy.
\label{tab:flux}
\end{table*}

\section{Radio galaxies in the cluster region }\label{sec:sources}

Figure 1 presents the GMRT 610 MHz contour image of PLCK171 at full resolution 
($\sim$5$^{\prime \prime}$), overlaid on the optical image
from the Digital Sky Survey (DSS). Discrete radio sources 
are labelled from S1 to S7. Their flux densities at 610 MHz
and positions, both measured from the image in Fig.~1, are summarized in 
Table \ref{tab:flux}. All have optical counterparts on the 
DSS, except for S7. No redshift measurements are currently
  reported for these galaxies in the literature and 
optical catalogs. The cluster X-ray centre (Table \ref{tab:PLCK171})
is approximately coincident with the peak of S4. Four sources 
(S1, S4, S5 and S6) show a narrow-angle tail (NAT) morphology. 
The tails of S4 and S6 are both oriented northward, while the tails 
of S1 and S5 point toward South-East and North-West, respectively, 
almost opposite to each other. Assuming that the 
NATs are at the redshift of the cluster, the length of their tails ranges, in
projection, from $\sim$90 kpc (S5) to $\sim$180 kpc (S6), and their
radio powers at 610 MHz are of the order of $\sim 10^{24-25}$ W
Hz$^{-1}$. These values are consistent with the 
range of size and radio power typically measured for NAT radio
galaxies \citep[e.g.,][]{2002ASSL..272..163F}.

The full resolution GMRT image at 235 MHz is presented in
Fig.~\ref{fig:235u} (grey scale), with the positions of sources 
S1 to S7 marked by red crosses and labels. Source S7 is not detected at 235
MHz. This is consistent with the fact that the source appears very
compact at 610 MHz, so it is probably an active AGN with flat or 
inverted spectrum. The 235 MHz flux 
densities of S1, S2, S5 and S6 are reported in Table \ref{tab:flux}, along with the spectral index
$\alpha$ computed between 235 MHz and 610 MHz. Sources S1 and
S5 are only partially detected at 235 MHz, thus the flux density 
and spectral indeces in Table \ref{tab:flux} should be considered as 
lower limits. Sources S3 and S4 appear blended together in Fig.~\ref{fig:235u}, 
therefore it is not possible to measure their individual flux density at 235 MHz. 
A total of 248.4 mJy is measured for the blend.

\begin{figure}
\centering
\includegraphics[width=8.5cm]{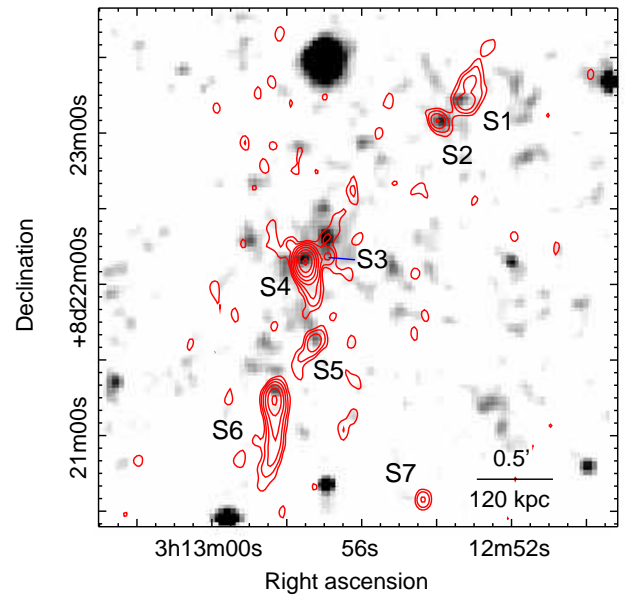}
\caption[]{GMRT full-resolution image at 610 MHz (contours) 
overlaid on the optical DSS image. The restoring beam is
$5.5^{\prime\prime}\times 4.5^{\prime\prime}$, in p.a. $0^{\circ}$.
The r.m.s. level in the image plane is 0.12 mJy beam$^{-1}$.
Contours start at 0.4 mJy beam$^{-1}$ and then scale by a factor of
2. No negative levels corresponding to the $-3\sigma$ level 
are present in this region. Labels indicate the discrete radio
sources, whose properties are summerized in Table 3.
The cluster X-ray center is at RA=3h12m57.4s, DEC=+08d22m10.3s (Tab.~\ref{tab:PLCK171}).}
\label{fig:610}
\end{figure}

\begin{figure}
\centering
\includegraphics[width=9cm]{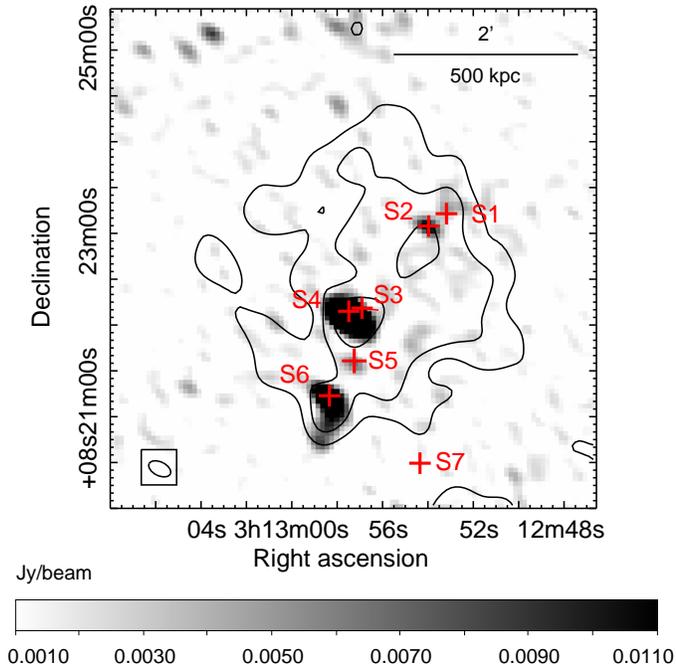}
\caption[]{GMRT full-resolution image at 235 MHz (gray scale) with
  overlaid the low-resolution 
image shown as contours. The restoring beam is
$15.4^{\prime\prime}\times 9.0^{\prime\prime}$, in p.a. $64^{\circ}$
(shown in the bottom-left corner)
and $28^{\prime\prime}\times 25^{\prime\prime}$, in p.a. $73^{\circ}$,
respectively.
The r.m.s. level is 0.8 mJy beam$^{-1}$ at full resolution and
2.3 mJy beam$^{-1}$ in the low-resolution image. Contour levels are 
6.9, 11.5, 18.4, 27.6 mJy beam$^{-1}$. No negative contours
corresponding to the $-3\sigma$ level 
are present in the portion of the image shown. The discrete sources 
detected at 610 MHz (Fig.~\ref{fig:610}; Table 3) are labelled and marked 
by the crosses.}
\label{fig:235u}
\end{figure}

\begin{table}[t]
\caption[]{Properties of the radio halo in PLCK171}
\begin{center}
\begin{tabular}{lc}
\hline\noalign{\smallskip}
\hline\noalign{\smallskip}
$S_{\rm 235 \, MHz}$  (mJy) &  $483\pm110$ \\
$S_{\rm 610 \, MHz}$ (mJy) & $>50$ \\
$S_{\rm 1400 \, MHz}$ (mJy) & $18\pm2$\\
$\alpha$ (${\rm 235 \, MHz \div 1400 \, MHz}$) & $1.84\pm0.14$\\
$P_{\rm 1400 \, MHz}$ (W Hz$^{-1}$) & $(4.9\pm0.1)\times10^{24}$\\
Linear size (Mpc)  & $\sim1$\\
\hline
\label{tab:halo}
\end{tabular}
\end{center}
\end{table}

\section{The giant radio halo}\label{sec:halo}

A low-resolution ($28^{\prime\prime}\times25^{\prime\prime}$) image at
235 MHz of PLCK171 is presented as contours in Fig.~\ref{fig:235u},
overlaid on the full-resolution image in gray scale. The image shows 
the residual emission after subtraction of the CLEAN
components of sources S1 to S6 from the $uv$ data. The CLEAN
components were obtained from an image produced using only 
the baselines longer than 3 k$\lambda$, containing information on
structures on angular scales $\lesssim 1.4^{\prime}$.
The presence of a large diffuse
source filling the central region of the cluster is evident. Its large linear size ($\sim 800$
kpc in Fig.~\ref{fig:235u}), roundish morphology and the central cluster location are consistent 
with the properties of a giant radio halo. 

We note that the peaks of the radio halo image in Fig.~\ref{fig:235u}
are approximately coincident with the position of some individual sources. To
estimate the possible residual contribution of these sources, we
made an image of the halo region from the source-subtracted data set,
using only the long baselines ($>3$ k$\lambda$). We indeed 
found some residual emission possibly associated with the blend of S3 and S4
($\sim 7$ mJy) and with S6 ($\sim 6$ mJy). This will be taken into
account when measuring the flux density and spectral index of the halo
in Sect.~\ref{sec:sp}.

\begin{figure*}
\centering
\includegraphics[width=16cm]{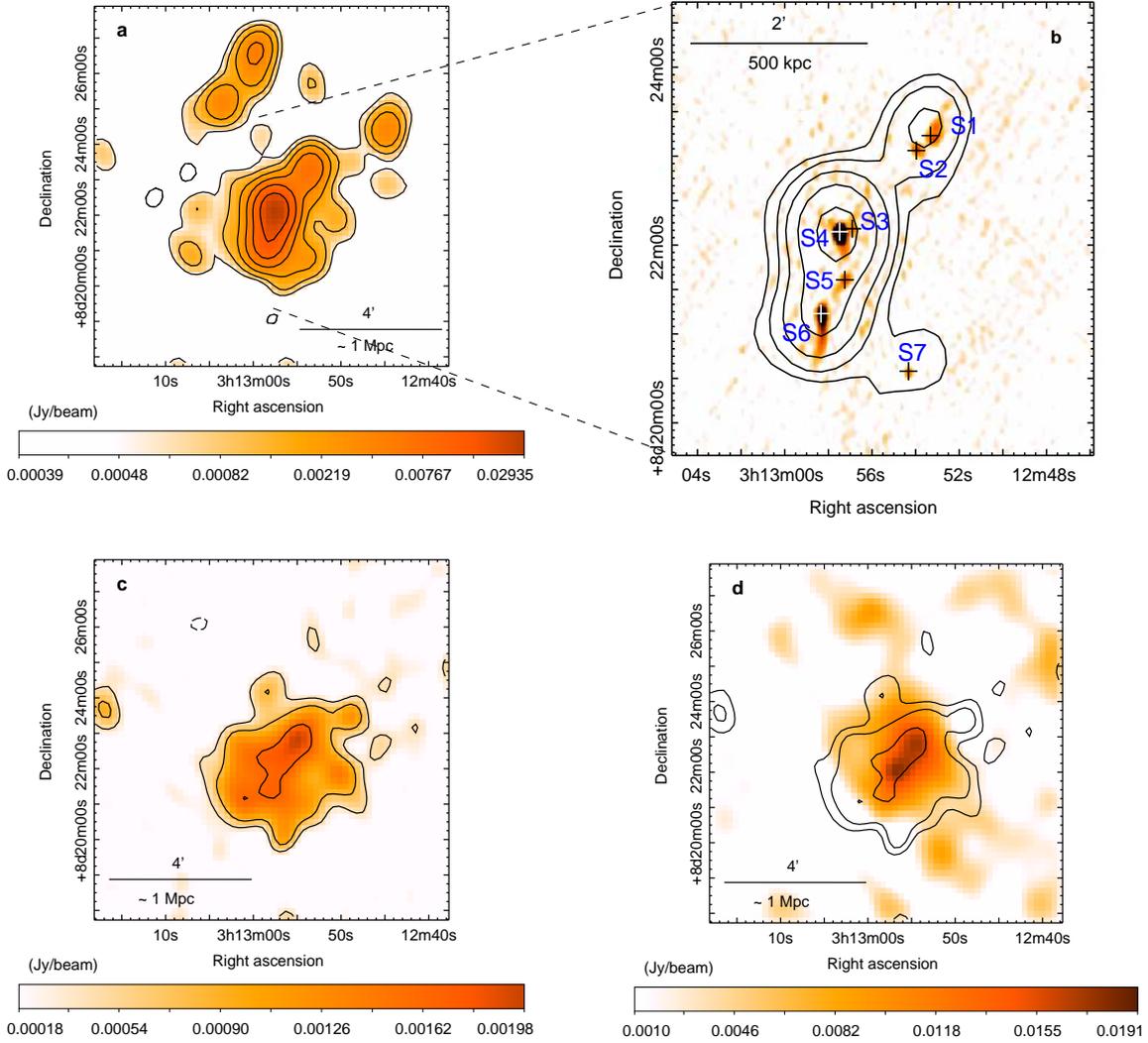}
\caption[]{VLA--D array images at 1.4 GHz of PLCK171 
obtained from the re-processed NVSS data.  
Contour levels are $-1$ (dashed), $1, 2, 4, 8, 16,... \times
3\sigma$. {\it a:} 
Contours and color scale image of PLCK171. The 
restoring beam is $58^{\prime\prime}\times 48^{\prime\prime}$, in p.a. $-5^{\circ}$
and $1\sigma$=0.17 mJy beam$^{-1}$.
{\it b:} Image obtained using only baselines
longer than 0.8 k$\lambda$. The restoring beam is $50^{\prime\prime}\times 
42^{\prime\prime}$, in p.a. $-7^{\circ}$ and
$1\sigma$=0.4 mJy beam$^{-1}$. Crosses and blue labels 
show the position of the discrete radio sources detected at 610 MHz 
(Fig.~\ref{fig:610}, Table 3). {\it c:} Image of the diffuse radio halo
after the subtraction of the discrete radio sources derived from 
the image in panel b. The image has been restored with a circular 
beam of $50^{\prime\prime}$. The r.m.s. noise is $1\sigma=0.12$ mJy beam$^{-1}$.
{\it d:} 1.4 GHz contours (same as panel c) overlaid on the 235 MHz
image of the radio halo convolved with a circular beam of $40^{\prime\prime}$.}
 \label{fig:nvss}
\end{figure*}

Hints of the halo are already visible on the public NVSS image at 1.4
GHz. To obtain a better quality image of the diffuse source, we re-analyzed 
the NVSS pointing (see Sect.~\ref{sec:vla}). Our 1.4 GHz images are
presented in Fig.~\ref{fig:nvss}. Panel a shows the total radio
emission, i.e. the halo and the radio galaxies.
Following the procedure adopted at 235 MHz, we re-derived
the image using only the baselines longer than 0.8 k$\lambda$ (angular scales $\lesssim 4^{\prime}$) to identify the contribution of
the discrete radio sources in the cluster area.
A central region of the resulting image is shown in panel b, with the position 
of the sources detected at 610 MHz (Fig.~\ref{fig:610}) highlighted by
crosses and labels. Even though the resolution of the two images is 
very different ($\sim5^{\prime \prime}$ at 610 MHz and $\sim50^{\prime \prime}$ at
1.4 GHz), there is a good match between the location of the
radio sources at 610 MHz and the structure imaged at 1.4 GHz.
The CLEAN components associated with this image were then
subtracted from the $uv$-data, and the resulting data set was 
used to obtain an image of just the radio halo, shown in 
Figure \ref{fig:nvss}c (colors and contours). An overlay of this 1.4
GHz diffuse emission on the GMRT 235 MHz image of the radio halo, convolved with a circular 
beam of $40^{\prime\prime}$, is shown in
Figure \ref{fig:nvss}d. The overall shape of the halo is similar at
both frequencies, while its extent is slightly larger at 1.4 GHz
($\sim 1$ Mpc). The difference is likely due to the low sensitivity and
quality of the 235 MHz observation, which was strongly affected by 
ionospheric scintillation during most of the observing time 
(Sect.~\ref{sec:gmrt}).

\begin{figure}
\centering
\includegraphics[width=8.5cm]{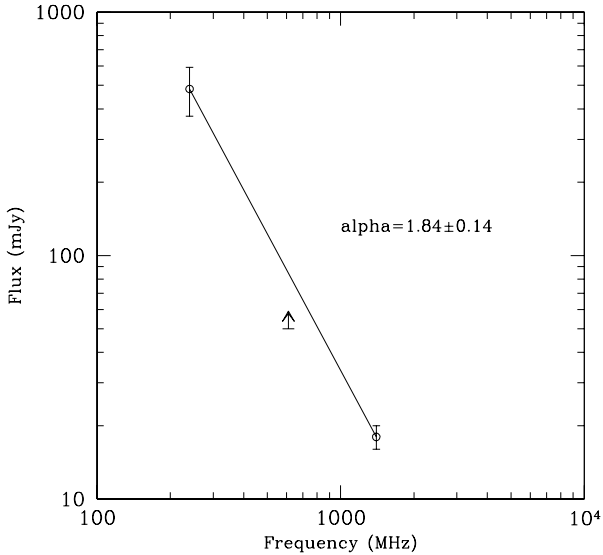}
\caption[]
{Spectrum of the radio halo between 235 MHz and 1.4
  GHz.}
\label{fig:sp}
\end{figure}

\begin{figure*}
\begin{center}
\includegraphics[width=16cm]{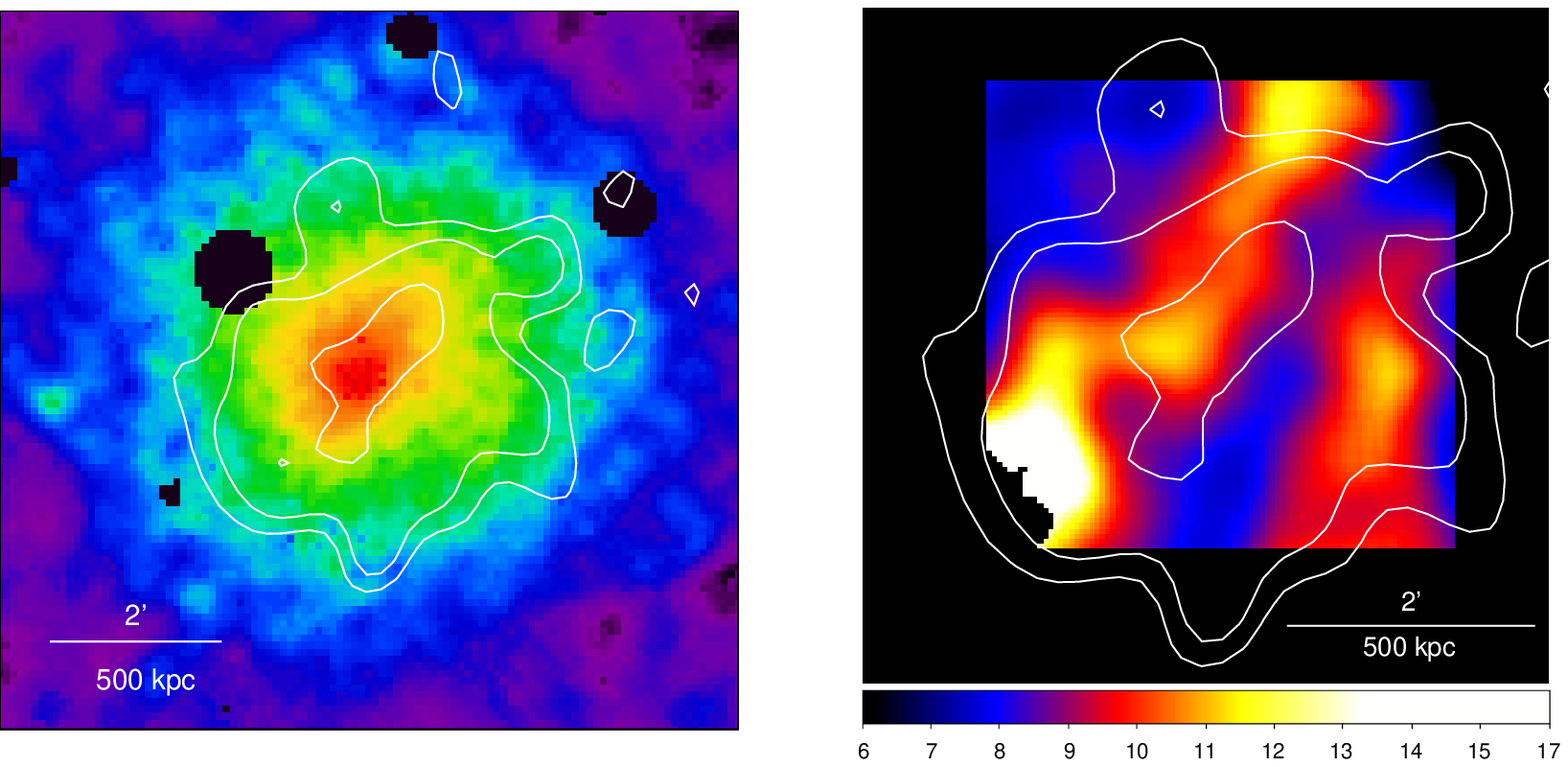}
\caption[]{{\em a:} Adaptively smoothed {\em XMM-Newton} image in the 0.4-7.2 keV band. 
{\em b:} Temperature map.  The colorbar displays the temperature in
keV. In both panels the 1.4 GHz image of the halo is overlaid as
contours (same as Fig. 3c).}
\end{center}
\label{fig:xray}
\end{figure*}

\section{Integrated spectrum of the radio halo}\label{sec:sp}

Using the images of the diffuse component alone (Fig.~\ref{fig:235u}
and Fig.~\ref{fig:nvss}c), we measured a total flux density of 
the halo of $483\pm110$ mJy at 235 MHz\footnote{The residual 13 mJy,
  possibly associated with the individual sources (see
  Sect.~\ref{sec:halo}), have been added into the error.}
and $18\pm2$ mJy at 1.4 GHz
in the region defined by the first contour in Fig.~\ref{fig:nvss}d
(3$\sigma$). 
These are also reported in Table \ref{tab:halo}, along with the other
main observational properties of the halo, and in Fig.~\ref{fig:sp}.
The derived total radio power at 1.4 GHz is $P_{\rm 1.4 \, GHz} = (4.9\pm0.1) \times 10^{24}$ W Hz$^{-1}$.

Our flux density values at 235 MHz and 1.4 GHz correspond to
a very steep spectral index, $\alpha=1.84\pm0.14$, which 
makes PLCK171 one of the steepest-spectrum giant halos known so far. 
However, large uncertainties affect our measurement. In particular,
it is possible that some of the halo flux density is missed in both
the reprocessed NVSS image and the GMRT 235 MHz image due to the 
short duration of the observations, which results in sparse coverages
of the $uv$ plane and loss of sensitivity to structure. Furthermore,
the low resolution of the 1.4 GHz image does not allow an accurate 
removal of the discrete radio galaxies in the halo region, and,
consequently, the halo flux density could be overestimated. 

A flux density of $\sim$45--50 mJy is measured from low-resolution ($\sim 30-40^{\prime\prime}$) images at 610 MHz 
(not shown here), obtained after the subtraction of the discrete sources.
However, these images detect only the brightest peaks of the halo 
and, therefore, its flux density at this frequency is only a lower
limit. The 610 MHz measurement is $\sim40\%$ lower than 
the value expected based on the spectral index between 235 MHz 
and 1.4 GHz (Fig.~\ref{fig:sp}). Our experience with the GMRT at 610 MHz suggests 
that the flux density losses at this frequency can indeed 
be conspicuous, up to $\sim$40$\%$ \citep{2008A&A...484..327V,2009ApJ...699.1288D,2010A&A...517A..43M}.

\section{X-ray analysis}

PLCK171 was observed with {\em XMM-Newton} for 14 ks in 2010 August
under ObsID 0656201101. The EPIC data were reduced using the Extended 
Source Analysis Software 
({\em XMM}-ESAS)\footnote{http://heasarc.gsfc.nasa.gov/docs/xmm/xmmhp\_xmmesas.html}
package part of SAS version 11.0.0. For the EPIC MOS detectors, the analysis of
galaxy clusters with this software was introduced by \citet{2008A&A...478..615S},
but the methodology has since been expanded to include the EPIC pn
data \citep[e.g.,][for a more
detailed description]{2012ApJ...747...32B}. 
In summary, a filtered event list is created by removing periods of excessive count
rate in the light curve caused by proton flaring events (retaining 13.7 ks,
13.4 ks, and 8.2 ks of the exposures for the MOS1, MOS2, and pn data, respectively), 
excluding anomalous MOS
CCDs (in this case CCDs 4, 5, and 6 for MOS1), and identifying and excluding
point sources.
Spectra and images are then extracted, including for the quiescent particle
background, which is derived from a database of filter wheel closed observations
that are matched to data from the unexposed corners of the chips.
The remaining background components (residual soft proton contamination,
instrumental lines, 
solar wind charge exchange (SWCX) emission, and the cosmic fore- and backgrounds
from the local hot bubble, the Galaxy, and extragalactic unresolved point sources)
are all explicitly modeled and determined empirically when fitting spectra in XSPEC.
While robust best fits of these components can be trickier to obtain in shorter
exposures such as this one, the smaller angular extent of this cluster provides
for a large local background region from which their parameters can be sufficiently
constrained.

In this case, nearly all of the cluster emission is taken to fall within a circle of
$320^{\prime\prime}$ radius; EPIC MOS and pn spectra from this region and
the remaining field of view out to $800^{\prime\prime}$
(containing only background) are simultaneously fit.
We find a global cluster temperature of $10.1 \pm 0.7$ keV, in good agreement
with that reported in Planck Collaboration et al. (2011c) 
of $10.65 \pm 0.42$ keV.
Our slightly lower temperature likely reflects a different source region and background
treatment, and our greater uncertainty is a consequence of treating background
components as free parameters.
For the image analysis, the best-fit values for the non-cosmic backgrounds 
(quiescent particle, soft proton, and SWCX)
are translated into images and subtracted from the data.

The resulting adaptively smoothed, full band (0.4-7.2 keV) image is presented in
Figure 5a with the 1.4 GHz radio halo contours overlaid
(same as Fig.~\ref{fig:nvss}c). Structurally, the central region is
elongated in the North--West/South--East direction with the 
X-ray peak slightly towards the SE side from the large-scale emission 
centroid. The radio halo and X-ray surface brightness generally follow each
other, as often observed in clusters hosting radio halos.
The asymmetric ICM morphology is consistent with a recent merger event.

To investigate the merger state in more detail, we constructed a temperature map
using smoothed narrow-band images fitted with the MeKaL plasma model,
using a method 
described in \citet{2000ApJ...541..542M}. 
Images are extracted in seven bands (0.4-0.8 keV, 0.8-1.1 keV, 1.1-1.35 keV,
1.9-2.1 keV, 2.35-3.2 keV, 3.2-4.8 keV, and 4.8-7.2 keV) that avoid instrumental lines.
A local background, accounting for Galactic and extragalactic non-cluster sources,
is estimated from the region outside the cluster in each image and
subtracted, after which the image is corrected by vignetting and
exposure maps.
Using the ESAS task {\tt comb}, which adjusts each exposure map to the
response of the MOS2, medium filter response, the three EPIC instruments are
combined to maximize the signal-to-noise before smoothing.
The images, with $2.5^{\prime\prime}$ pixels, are variably smoothed by a 
10 pixel Gaussian kernel at the image edges, which transitions to a $\sim 5$ pixel 
kernel inversely following the wide band surface brightness to the 0.2 power.
Chip and point source gaps are interpolated over (for purely cosmetic purposes).
Then, an absorbed MeKaL model with fixed absorption ($2.8\times10^{21}$ cm$^{-2}$),
abundance (0.3 solar), and redshift (0.27)
is fit to the coarse spectrum in the seven energy bins corresponding to each pixel.
The resulting best-fit temperatures are shown in Figure 5b, 
which have 1-sigma errors on the order of 1 keV 
in the center and $\sim 2$ keV near the edge.

The general picture that emerges is a roughly isothermal central
region, without a central cool core and with a slight, but only
marginally significant, temperature peak ($\sim 11$ keV) coincident with the X-ray peak.
The temperature peak extends in the same direction, 
but more narrowly, as does the broad band surface brightness. 
This elongation follows that evident in the core of the radio halo,
suggesting a merger along the NW-SE axis.
The temperature of the cooler region to the SW of the X-ray peak
($\sim 7-8$ keV) deviates 
from the temperatures along the NW-SE axis at the 2.3$\sigma$ level and is
confirmed in direct spectral fits.
Deeper X-ray observations are necessary to discern exactly what this feature
is and its relationship to the merger.

\begin{figure}
\centering
\includegraphics[width=8cm]{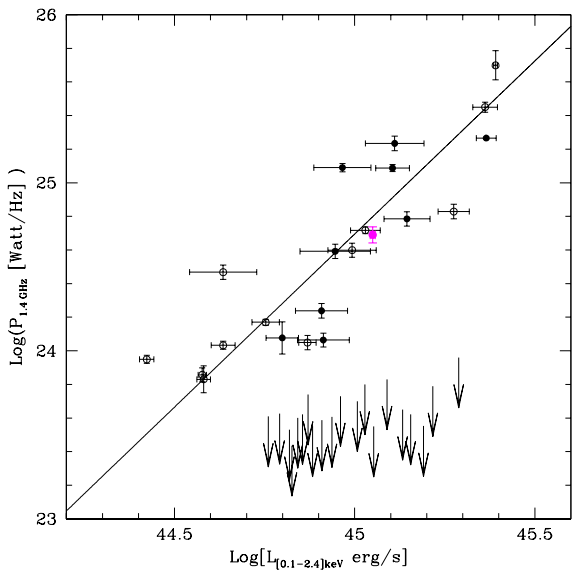}
\caption[]{Distribution of the radio-halo clusters in the
$P_{\rm 1.4 \,GHz}-–L_X$ plane from Brunetti et al. (2009). 
Filled black symbols are the GMRT radio-halo clusters
and open black symbols are other radio-halo clusters from 
the literature. The magenta point marks the position of PLCK171,
using the cluster X-ray luminosity estimated within $R_{500}$ (Planck
Collaboration et al. 2011c) and the 1.4 GHz radio power in Table 4.
The solid line is the best fit to the distribution of giant radio
halos from Brunetti et al. (2009).}
\label{fig:lxlrad}
\end{figure}

\begin{figure}
\centering
\includegraphics[width=8.5cm]{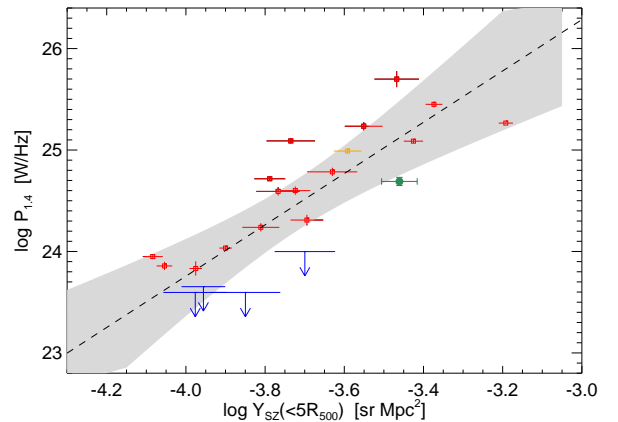}
\caption[]{Distribution of the radio-halo clusters in the
$P_{\rm 1.4 \,GHz}-–Y_{SZ}(<5R_{500})$ sr Mpc$^2$ plane (Basu 2012). The green
symbol marks the position of PLCK171. The red symbols are the known 
radio-halo clusters, the yellow symbol is the mini-halo in A\,2390 and 
blue symbols are the upper limits from the GMRT radio-halo survey
(Brunetti et al. 2007).}
\label{fig:psz}
\end{figure}

\begin{figure}
\centering
\includegraphics[width=8.5cm]{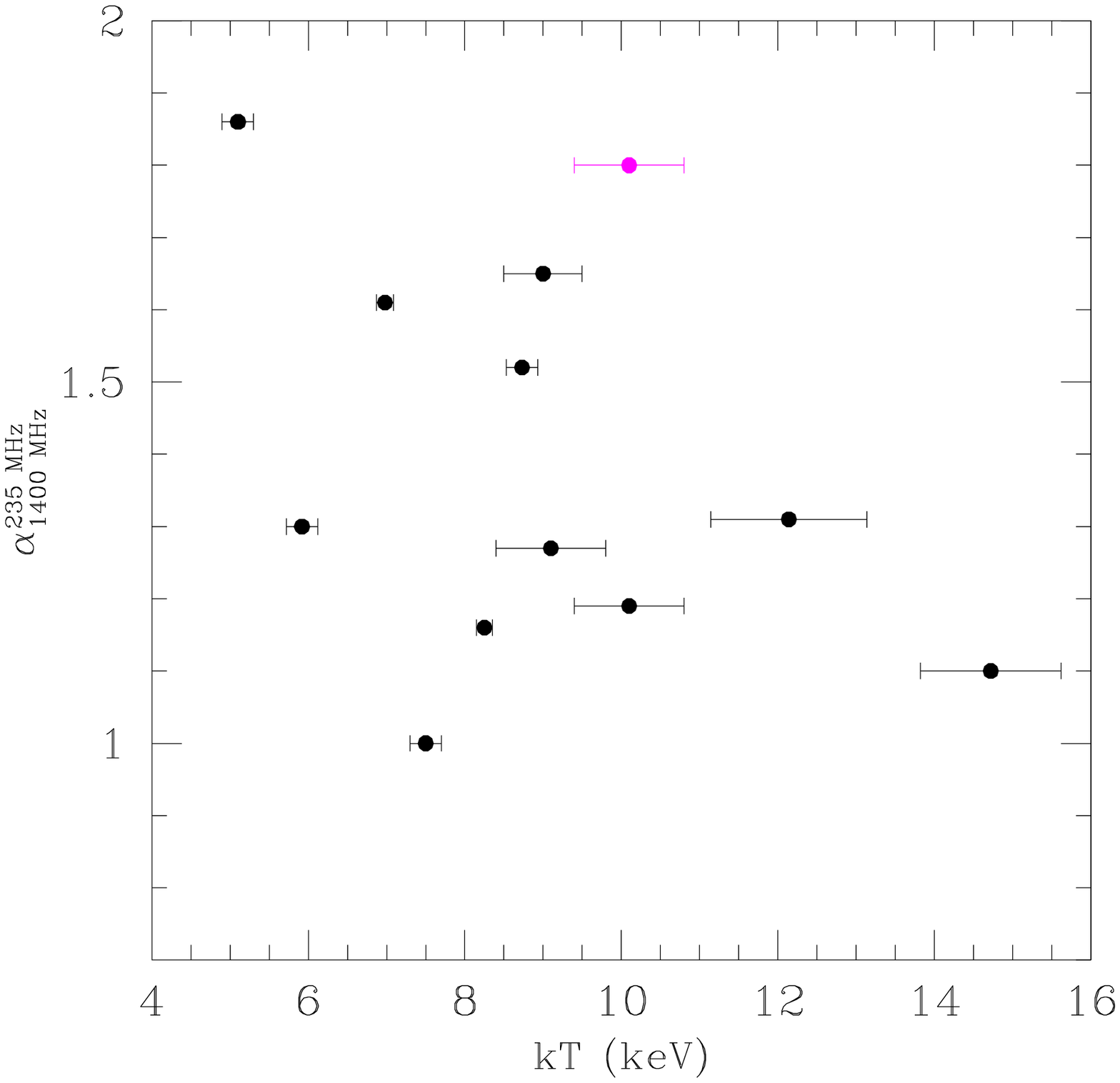}
\caption[]{Spectral index of radio halos in the 235 MHz-1.4 GHz
  interval as function of the cluster X-ray temperature (adapted from Venturi
  et al. 2013). The magenta points marks the position of PLCK171.}
\label{fig:talpha}
\end{figure}

\section{Discussion}

The {\em Planck} mission has discovered several new clusters using the SZ
signal. These systems were not detected from the all sky X-ray
surveys, such as the ROSAT All Sky Survey, because of their low 
surface brightness and being near the flux limit of the surveys.
The newly discovered {\em Planck} clusters are mostly
massive systems with highly disturbed ICM. With the current knowledge of the
connection between cluster mergers and the presence of giant radio halos
\citep[][and references therein; see also \cite{2011A&A...532A.123R}]{2010ApJ...721L..82C}, these
new clusters are excellent targets for searching for new
halos. In this paper, we reported the discovery of the first giant
radio halo in these new {\em Planck} clusters -- in PLCK171 -- based 
on GMRT observations at 235 MHz and 610 MHz and re-analysis of the VLA 1.4 GHz NVSS data.

The diffuse halo is $\sim$1 Mpc in extent and is cospatial with the
brightest X-ray emission, as typically observed in radio-halo clusters 
\citep[e.g.,][]{2004ApJ...605..695G,2012arXiv1210.7617V}.
The spectral index of
the halo between 235 MHz and 1400 MHz is quite steep,
$\alpha=1.84\pm0.14$. This is one of the steepest slopes measured 
for a giant radio halo so far -- in fact, it is similar to $\alpha\sim1.9$ of the
prototype USSRH in A\,521 \citep[][]{2009ApJ...699.1288D}.
However, the spectral index
of the PLCK171 halo is affected by large uncertainties. In
particular, due to the short usable time of our 235 MHz observation 
and short integration time of the NVSS pointing, it is possible that 
the halo extends more than imaged here, and a fraction of its flux density 
is missed at both frequencies. Moreover, the low resolution 
of the 1.4 GHz image does not allow an accurate subtraction of the
discrete radio galaxies enclosed in the halo emission. Therefore, 
deeper observations at 1.4 GHz with higher angular resolution, 
and deeper exposures at lower frequencies are essential before any conclusive
interpretation of the spectral index
of this halo can be made. If future observations confirm its
steep radio spectrum, PLCK171 would be another important case of
USSRH (of which only a few other examples are known so far; e.g., 
Brunetti et al. 2008, Macario et al. 2010, van Weeren et al. 2011, and
a candidate in Giacintucci et al. 2011), whose study is 
crucial to understand the physics behind the origin of radio halos in 
general. While such objects are expected in turbulent reacceleration models, the
existence of halos with $\alpha >1.5$ poses serious problems for
secondary models due to the required large energy in cosmic
ray protons (Brunetti et al. 2008, Macario et al. 2010, van
Weeren et al. 2011). Such energies are above the current upper limits
from the $\gamma$-ray observations of nearby clusters (Aharonian et al. 2009, 
Ackermann et al. 2010,  Jeltema \& Profumo 2011), assuming
the $\mu$G cluster magnetic field values indicated by Farady rotation
measure studies (e.g., Bonafede et al. 2010 and references therein).

The analysis of the X-ray {\em XMM-Newton} observation of PLCK171 
reveals high temperature ($\sim 10$ keV) and 
signatures of a recent merger in both the surface brightness and
gas temperature distributions. A NW-SE elongation of the central X-ray
emission, coupled with a similar structure in the temperature map,
suggests that the merger occurred along this axis. The radio halo 
displays a similar asymmetry in its central region, suggesting a
correlation between the thermal and non-thermal emissions, as typically
observed in radio-halo clusters.

Figure \ref{fig:lxlrad} shows the distribution of the clusters of the 
GMRT radio-halo survey (Venturi et al. 2007, 2008), along
with other radio-halo clusters from the literature, in the $P_{\rm
  1.4 \,GHz}$--$L_X$ plane (from Brunetti et al. 2009).  The
bimodality evident in this diagram (with some clusters lacking diffuse
radio emission, while those that exhibit radio halos follow a correlation)
is one of the main 
outcomes of the GMRT radio-halo survey and provides quantitative 
support to reacceleration models for the origin of radio halos
(Brunetti et al. 2007, 2009). The separation between
clusters hosting a radio halo and those without (upper limits) is 
most likely caused by the different dynamical properties of the clusters: 
clusters without a radio halo tend to be more relaxed
systems, while radio halos are exclusively found in disturbed 
clusters (Cassano et al. 2010).

PLCK171 (magenta symbol) falls on the radio halo correlation. Furthermore,
our X-ray analysis shows that the cluster is unrelaxed, as expected 
for a system hosting a radio halo. Ongoing dynamical activity in PLCK171 
is also suggested by the presence of four NATs in its central Mpc 
region, with tails pointing away from the cluster center, suggesting
infall of multiple subctructures onto the main cluster and bulk motions 
of the ICM driven by the ongoing merger (e.g., Bliton et al. 1998).
Thus, the radio halo in PLCK171 
provides further support to the existence of a direct link
between the radio halo phenomenon and cluster mergers. 

A scaling relation between the radio power at 1.4 GHz and the
integrated SZ effect measurement (Y) has been recently presented by
Basu (2012). Such a relation is a direct probe of the connection between
the mass of the cluster and presence of a radio halo. In
Fig.~\ref{fig:psz} we report the position PLCK171 in the $P_{\rm 1.4
  \,GHz}$-$Y$ plane using the integrated Y published
in \cite{2011A&A...536A...9P}.
PLCK171 (green square)
follows the scaling relation obtained for the other radio-halo
clusters. 

What is unusual about PLCK171 is its combination of
high gas temperature ($\sim 10$ keV) and ultra-steep 
spectrum of its radio halo ($\alpha \sim 1.8$, with the caveats given above). 
If the steep spectrum is confirmed, it would be the hottest cluster 
with a USSRH -- all other USSRHs have $kT\sim5-9$ keV, as visible in Fig.~\ref{fig:talpha}, that reports 
the distribution of radio-halo clusters in the $\alpha_{\rm 1400 \, MHz}^{\rm 235 \,
  MHz}-kT$ plane adapted from Venturi et al. (2013). The magenta point
marks the position of PLCK171 using $kT=10.1\pm0.7$ keV, as measured in
Sect.~6. Thus, this object may provide an important data point for 
our understanding of giant radio halos in clusters.

\section{Conclusions}
 We have discovered 
a $\sim1$ Mpc-size radio halo in PLCK171, using GMRT observations at 235 MHz and 610 MHz
and re-analysis of NVSS 1.4 GHz data. This is the first
giant radio halo found in the new clusters 
discovered by {\em Planck}. 

With a spectral index $\alpha \approx 1.8$, 
this source might be another system with a very 
steep radio spectrum, so far found in only a few other clusters.
However, 
deeper 
multi-frequency observations are required before any definitive
interpretation of the spectral index of this radio halo.

The X-ray luminosity of the cluster and radio power of the halo 
at 1.4 GHz are consistent with the correlation known 
to be followed by radio-halo clusters. Its location in the 
$P_{\rm 1.4   \,GHz}$-$Y$ plane also follows the scaling relation 
found for clusters with radio halos. The asymmetries observed 
in the X-ray surface brightness and temperaure distributions,
derived from the {\em XMM} data, 
indicate a disturbed ICM resulting from a recent merger.
This is in line with the expectation that giant radio halos are 
exclusively associated with cluster mergers. 

With an X-ray luminosity of 1.13$\times10^{45}$ erg s$^{-1}$  and a
redshift of $z=0.27$, PLCK171 represents the kind of clusters that have been
missed in X-ray flux limited surveys, such as the Rosat All Sky Survey, and, consequently, in radio
surveys based on X-ray selection criteria, such as the GMRT radio-halo
survey (Venturi et al. 2007, 2008). It is therefore important to consider biases due to clusters
such as PLCK171 in X-ray flux limited samples selected for radio surveys.

\acknowledgements

We are deeply grateful to Rossella Cassano for useful comments and
suggestions, and for providing Fig.~6. We thank Kaustuv Basu for
kindly providing Fig.~7. We thank the staff of the 
GMRT for their help during the observations.
GMRT is run by the National Centre for Radio Astrophysics of the Tata
Institute of Fundamental Research. The National Radio Astronomy Observatory is a facility of the National Science 
Foundation operated under cooperative agreement by 
Associated Universities, Inc.
SG acknowledges the support of NASA through Einstein Postdoctoral
Fellowship PF0-110071 awarded by the Chandra
X-ray Center (CXC), which is operated by SAO. This research was
supported by an 
appointment to the NASA Postdoctoral
Program at the Goddard Space Flight Center, administered by Oak Ridge
Associated Universities through a contract with NASA.

{\em Note added in proof}. This X-ray source was in fact detected by ROSAT
as 1RXS031257.2+082236, but not included in any ROSAT cluster catalogs. We thank
Alastair Edge for pointing out the ROSAT detection of this source.

\end{document}